\documentclass[journal]{IEEEtran}

\def\Journal#1#2#3#4{{#1} {\bf #2}, #3 (#4)}
\def\NIMA{{Nucl. Instrum. Methods} A}
\def\PLB{{Phys. Lett.}  B}
\def\JINST{JINST}
\def\JHEP{JHEP}

\usepackage{atlasphysics}

\ifCLASSINFOpdf
  \usepackage[pdftex]{graphicx}
  \graphicspath{{}}
  \DeclareGraphicsExtensions{.pdf,.jpg,.png}
\else

\fi

\usepackage{lineno}

\begin{document}
\pagenumbering{gobble}

\title{Upgrade of the ATLAS Liquid Argon Calorimeters for the High-Luminosity LHC}

\author{T.G. McCarthy$^{\dagger}$ on behalf of the ATLAS Liquid Argon Calorimeter Group
 \\ \ \\  \textit{$^{\dagger}$Max-Planck-Institut f$\ddot{\textrm{u}}$r Physik} \\ \textit{F$\ddot{\textrm{o}}$hringer Ring 6, 80805 M$\ddot{\textrm{u}}$nchen, Germany} \\ \textit{e-mail: tmccarth@mpp.mpg.de} \\ \textit{www.mpp.mpg.de}}

\maketitle

\begin{abstract}
The increased particle flux at the high luminosity phase of the Large Hadron Collider (HL-LHC), with instantaneous luminosities of up to 7.5 times the original design value, will have an impact on many sub-systems of the ATLAS detector.  This contribution highlights the particular impacts on the ATLAS liquid argon calorimeter system, together with an overview of the various upgrade plans leading up to the \mbox{HL-LHC}.  The higher luminosities are of particular importance for the forward calorimeters (FCal), where the expected increase in the ionization load poses a number of problems that can degrade the FCal performance such as beam heating and space-charge effects in the liquid argon gaps and high-voltage drop due to increased current drawn over the current-limiting resistors.  A proposed FCal replacement as a way to counter some of these problems is weighed against the risks associated with the replacement.  To further mitigate the effects of increased pile-up, the installation of a high-granularity timing detector at the front face of each end-cap cryostat is also currently under consideration.  Several different sensor technologies and layouts are being investigated.
\end{abstract}

\IEEEpeerreviewmaketitle

\section{Introduction}\label{sec:intro}
\IEEEPARstart{T}{he} ATLAS Experiment \cite{atlas-paper} at CERN's Large Hadron Collider (LHC) has recently entered a year-end shutdown period, roughly marking the halfway point in its second data-taking run. By mid-2016 ATLAS was collecting data from proton-proton collisions at a centre-of-mass energy of \mbox{$\rts=13\,$\TeV} at both the design proton-proton bunch-spacing of $25\,$ns and at instantaneous luminosities reaching the design value of \mbox{$\mathcal{L} = $\highL}.  The end of Run-2 at the end of 2018 will be followed by an extended shutdown period from 2019-2020 which will feature a number of \textbf{Phase-I} upgrades to the detector \cite{atlas-phase1-tdr}.  A further long-term shutdown period from 2024-2026 will feature a series of \textbf{Phase-II} upgrades \cite{atlas-phase2-scoping-document}.  The data-taking period beyond the Phase-II upgrades is referred to as the High-Luminosity LHC (\mbox{HL-LHC}), during which the instantaneous luminosities will be expected to climb to roughly 7.5 times their design value.

Such high instantaneous luminosities have consequences on various aspects of the experiment, notably the trigger and data storage rates, the degradation of electronics and other hardware components more sensitive to the increased radiation and particle fluxes, and the potential loss in sensitivity of physics measurements as a result of pile-up: a measure of the increase in noise as a result of additional soft proton-proton interactions other than the hard-scatter interaction of interest.  A useful number for quantifying the amount of pile-up during proton-proton data-taking is the average number of interactions per bunch-crossing, $\aipbc$.  Over the course of the \mbox{$\rts=7\,\TeV$} data-taking period in 2011 $\aipbc$ had an average value of approximately 9, while at the \mbox{HL-LHC} it is expected to reach values in the range of 140-200.

\begin{figure}[h]
\centering
\includegraphics[width=3.5in]{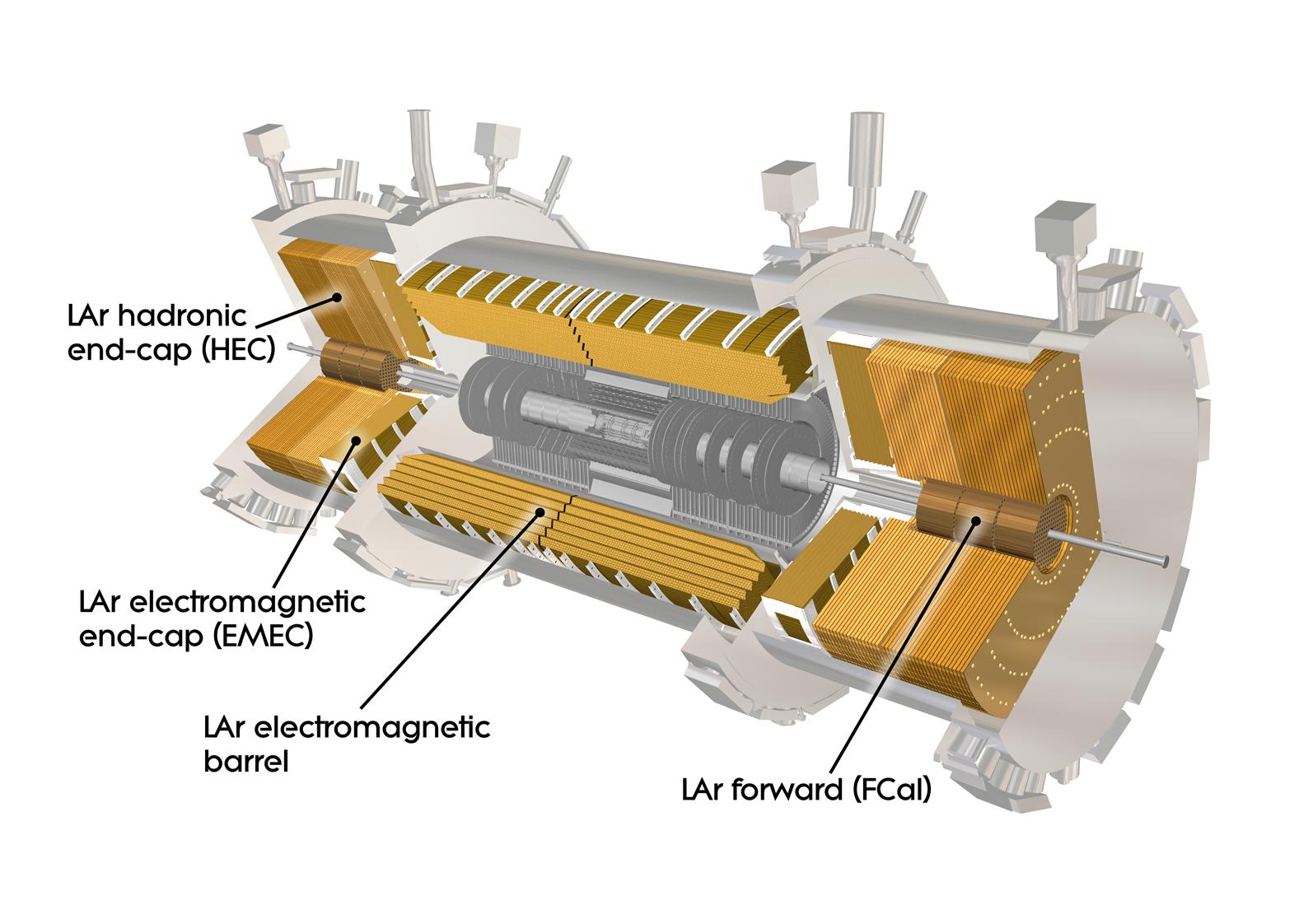}
\caption{Cut-away view of the ATLAS liquid argon calorimeter system \cite{atlas-paper}.}
\label{fig:atlas-lar}
\end{figure}

The liquid argon (LAr) calorimeter system, shown in Figure~\ref{fig:atlas-lar}, will also be affected by the conditions of the \mbox{HL-LHC}.  The combined LAr calorimeters provide electromagnetic and hadronic coverage within the pseudorapidity region $|\eta|\leq4.9$ and provide a total of approximately 182,500 individual readout channels or cells \cite{atlas-lar-tdr}.  The various components are housed in three separate cryostats (one barrel and two end-cap) operating at liquid-argon temperatures.  The LAr system consists of sampling calorimeters where liquid argon serves as the active material.  The absorber materials differ for the various sub-detector regions: lead is used in the electromagnetic barrel and end-cap calorimeters, copper in the hadronic end-cap, and both copper and tungsten in the forward region. The LAr calorimeters were originally designed for a 10-year operation period with a total integrated luminosity on the order of \mbox{$\int \mathcal{L}\textrm{d}t = 1000\,$\ifb}.

The planned LAr upgrades, both for Phase-I and Phase-II, are driven by several aspects: the need for better pile-up suppression in order to be able to retain high signal efficiencies in the context of limited trigger rates; the ability for the hardware to withstand higher than nominal radiation doses, due to a foreseen increase in the final integrated luminosity over the lifetime of the experiment by a factor of 3; the ageing and activation of the present detector components; and the available of faster electronics since the original LAr calorimeter design and construction.


\section{Trigger and Readout Electronics}\label{sec:trigger-electronics}

The present LAr calorimeter level-1 triggering system features a rather course readout granularity in the form of so-called trigger towers (TT) -- sums of the analog signal pulses from all readout channels within a fixed \mbox{$\Delta \eta \times \Delta \phi = 0.1 \times 0.1$} area.  The TT sums are also performed over all longitudinal calorimeter layers.  The level-1 (L1) trigger acceptance rates during Run 3 (immediately follwing the Phase-I upgrades) will continue to be limited to 100 kHz with a latency of \mbox{$3\,\mu$s}.  The increased pile-up during Run 3 (with expected values \mbox{$60\leq\aipbc\leq80$}) would saturate such L1 trigger rate limits.  

\begin{figure}[h]
\centering
\includegraphics[width=3.25in]{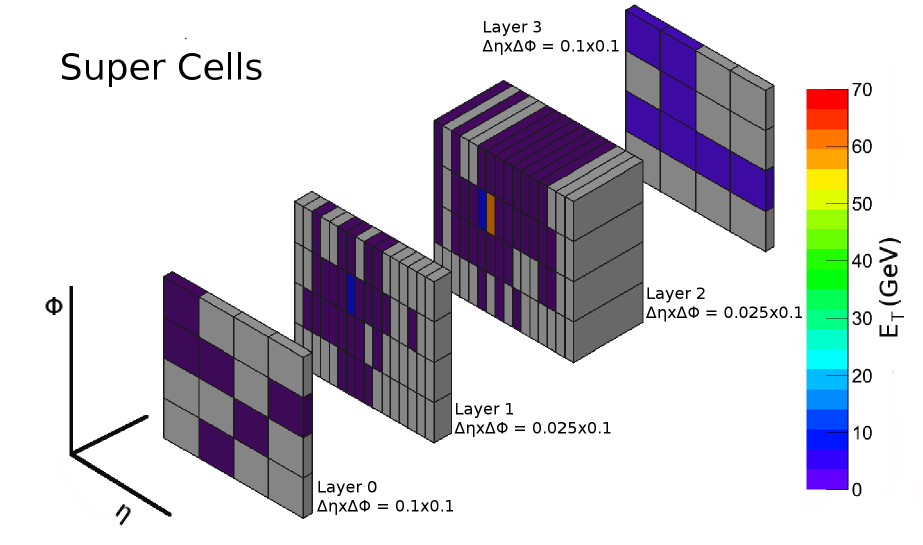}
\caption{The energy deposits of a trigger-level electron candidate with a transverse momentum of 70 \GeV \ based on the enhanced Super Cell granularity to be incorporated as a part of the the Phase-I LAr upgrades \cite{atlas-phase1-tdr}.}
\label{fig:trigger-super-cells}
\end{figure}

In order to avoid any significant loss of signal efficiency by raising \pT \ thresholds for the various triggers, the major LAr upgrade project for Phase-I involves the incorporation of so-called Super Cells (SC) at the L1 trigger level.  In addition to reading out each layer separately for the SCs as opposed to performing the layer sum as is currently done with the TTs, the readout granularity of the central two layers will be reduced to \mbox{$\Delta \eta \times \Delta \phi = 0.025 \times 0.1$}.  For the hadronic end-cap calorimeter the layer sum will continue to be performed.  What is currently a single trigger tower in the barrel region of the calorimeter will correspond to 10 Super Cells following Phase-I.  Figure~\ref{fig:trigger-super-cells} shows the energy deposits in 160 SCs resulting from a single electron with transverse momentum 70 \GeV \ based on simulation.  The same 160 SCs would be read out as 16 TTs in the present L1 configuration.
   
\begin{figure}[h]
\centering
\includegraphics[width=3.0in]{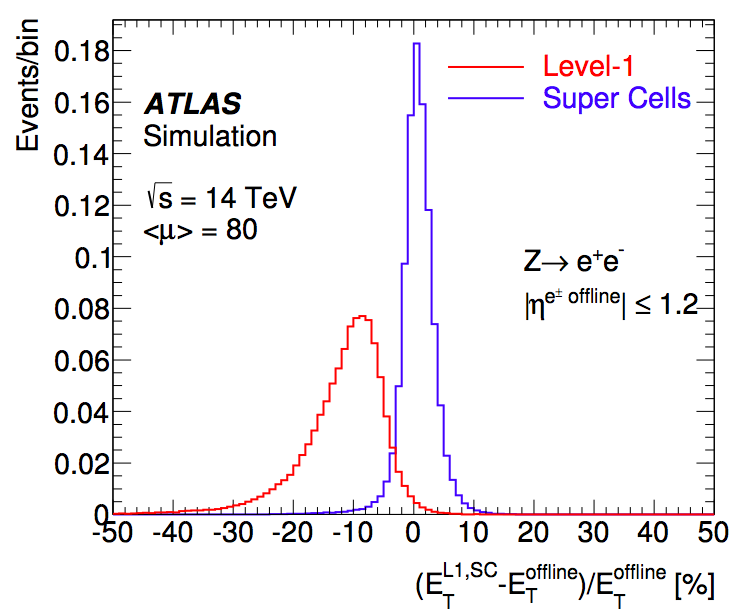}
\caption{The difference between L1 trigger-level and offline reconstructed energies of electron candidates from simulated $\Zee$ events.  The results based on the present L1 trigger tower energies (red line) can be compared with those employing the Super Cell granularity (blue line) \cite{atlas-phase1-tdr}.}
\label{fig:trigger-et-reso}
\end{figure}

The increased SC granularity allows a number of discriminating longitudinal and lateral shower shape variables to be defined.  Such variables can then be exploited to bring the L1 trigger rates to tolerable levels by applying cuts which target the pile-up backgrounds without adversely impacting the signal efficiencies.  Improved $e/\gamma$ identification will be possible by reducing the jet backgrounds for EM triggers.  The energy response and resolution using SCs will also significantly improve as highlighted in Figure~\ref{fig:trigger-et-reso}.

For Phase-II a replacement of all major front-end and back-end electronics will be performed.  These hardware upgrades will accommodate the increases to the L0/L1 trigger rates and latency, and will further avoid any complications arising from the increased ageing and radiation exposure of the various components which were designed to tolerate a total integrated luminosity on the order of \mbox{$\int \mathcal{L}\textrm{d}t = 1000\,$\ifb}, a factor of three lower than the present anticipated value.  Following the Phase-II upgrades, the full granularity will be incorporated into a new L0 trigger.  It should be noted that the Phase-I trigger upgrades will require forward compatibility with the Phase-II upgrades.

The details concerning the front-end and back-end electronics were the focus of another contribution, but additional information is available from references \cite{atlas-phase2-scoping-document,atlas-lar-back-end-electronics,atlas-lar-front-end-electronics}.

The ability to continue during Run 3 and at the \mbox{HL-LHC} with low trigger \pT \ thresholds is of paramount importance to the discovery potential for new physics at ATLAS.  The Phase-I and Phase-II upgrades to the L0 and L1 LAr trigger system will greatly benefit ATLAS in future searches for new physics, as well as precision Standard Model measurements.

\section{Forward Calorimeters}\label{sec:fcal}

\begin{figure}[h]
\centering
\includegraphics[width=2.5in]{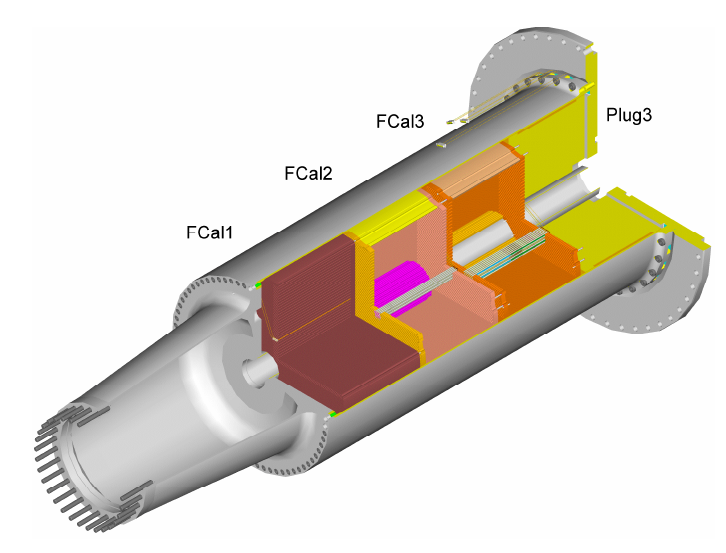}
\caption{Cut-away view of one of the FCal units, which consists of three modules together with a passive brass plug \cite{atlas-fcal-paper}.}
\label{fig:fcal-modules}
\end{figure}

The far-forward regions in ATLAS feature the forward calorimeters (FCal) \cite{atlas-fcal-paper} which cover the pseudorapidity range \mbox{$3.2\leq|\eta|\leq4.9$}.  The FCal absorber materials both allow for the initiation of electromagnetic and hadronic showers as well as provide containment for such showers in the high-radiation region close to the LHC beampipe.  An energy measurement is made possible via the interactions of high-energy ionizing particles traversing regions of the calorimeter containing the active material.  Each of the two FCal units consists of three longitudinal segmentations (FCal1, 2 and 3), depicted in Figure~\ref{fig:fcal-modules}.  In all three modules LAr acts as the active material.  The absorbing or inactive materials in the modules differ: copper is used in FCal1 which is referred to as an electromagnetic calorimeter, whereas tungsten is primarily used for FCal2 and 3, which are referred to as hadronic calorimeters.  The names reflect the importance of the calorimeters in their ability to contain electromagnetic and hadronic showers, respectively.

Individual FCal electrodes are arranged parallel to the beampipe at fixed separations in a hexagonal structure and consist of a central rod surrounded by a concentric copper tube.  Between the two a helically wound PEEK fibre maintains a very thin gap containing the active LAr.  Ionization electrons are released in this gap when high-energy particles traverse the liquid argon.  By applying a large potential difference across the LAr gap, the ionization electrons can be collected at the central rods and a signal read out in the form of a current pulse.  The individual signals are read out in groups, with FCal1 offering the finest granularity. A total of approximately 3500 individual readout channels provide the energy measurements in the combined FCal system.

The narrow LAr gaps, which range from approximately \mbox{$250\,\mu$m} in FCal1 to \mbox{$500\,\mu$m} in FCal3, allow for a fast signal readout together with a small level of ion buildup.  

The forward calorimeters are particularly important for ATLAS in the context of missing transverse momentum (\MET) measurement and in order to identify signal processes which feature high-energy, forward-going objects, such as the quark-initiated hadronic jets prominent in vector-boson fusion and vector-boson scattering events (VBF and VBS).  The importance of physics in the forward region can be underlined given that a number of beyond the Standard Model (BSM) models predict the existence of as-of-yet unobserved, heavy resonances which could be produced at the LHC via VBF.

\subsection{Anticipated Challenges for the FCal at HL-LHC}\label{sec:problems-hllhc}

At the instantaneous luminosities and particle fluxes expected at \mbox{HL-LHC}, the FCal will face a number of challenges, despite the radiation hardness of the active and absorber materials themselves.  

The conditions at the \mbox{HL-LHC} will lead to an energy deposition in FCal1 of roughly 28 \TeV \ per bunch crossing.  Since the temperature window for the argon to remain in a liquid state is small, this increased heat, which can be translated to roughly 110 W, was initially thought to pose a danger to the FCal.

A detailed study was carried out to assess the risk of liquid argon boiling during \mbox{HL-LHC} running conditions.  Predictions from a finite element analysis simulation were first compared to temperature measurements from previous \mbox{$\rts=8$} and \mbox{$\rts=13\,\TeV$} LHC runs.  These measurements were made possible via a number of sensors situated at various positions within the FCal units and sensitive to temperature changes on the order of \mbox{$\Delta T= 10\,$mK}.  

Once a good level of agreement between data and simulation was established, the simulation was used to predict the expected heat dissipation and temperature variations in the FCal during \mbox{HL-LHC} data-taking conditions.  The maximum expected temperature increase at the \mbox{HL-LHC} based on the current method of regulating the LAr temperature was found to be \mbox{$\Delta T_{\textrm{\small{max,HL-LHC}}} = 2.8^{+0.7}_{-0.9}\,K$}.  The nominal LAr temperature is \mbox{$T = 88.35\,$K}.  Only with an upwards deviation of \mbox{$\Delta T_{\textrm{\small{max,HL-LHC}}}$} at the 3-$\sigma$ level would a temperature of \mbox{$93.1\,$K} be reached in the region of the FCal most sensitive to the increased heat -- the temperature at which the liquid argon is expected to begin to undergo a phase transtion.

The results of the study therefore confirm that at the \mbox{HL-LHC} the current temperature regulation scheme should prove adequate to avoid any LAr bubble formation.  The temperature will however continue to be monitored closely, and the ability to reduce the LAr temperature by up to \mbox{$\Delta T=1.6\,$K} in order to remain well within the required safety limits, should the need to do so arise, was also demonstrated in a test involving one of the end-cap cryostats.

In addition to the beam heating effects, the increased particle fluxes at the \mbox{HL-LHC} will result in elevated ion build-up in the active LAr gaps, an effect which degrades and distorts the electric fields and leads to signal peaks with lower amplitudes.  This will be exacerbated by a high-voltage sagging effect due to the creation of large currents through the calorimeter's protection resistors.  Both of these ultimately translate to a degradation of the signal-to-noise level and to larger uncertainties in the energy measurement.

\subsection{Proposed Finer-Granularity FCal Replacement}\label{sec:fcal-sfcal}

With these potential limitations and issues in mind, a proposal was made to replace the FCal as a part of the Phase-II upgrades with an improved so-called sFCal which would feature both the same basic design and an identical choice for both active and passive materials as in the original FCal, but with a number of notable improvements \cite{atlas-phase2-scoping-document}.

Although small values were chosen for the LAr gaps in the original FCal design, they will prove too large at the high instantaneous luminosity levels present at the \mbox{HL-LHC}.  The values of the LAr gaps for the improved detector were therefore chosen to be roughly \mbox{$100\,\mu$m}, \mbox{$200\,\mu$m} and \mbox{$300\,\mu$m} for sFCal1, 2 and 3, respectively -- reductions by factors of 2.7, 1.9 and 1.7 relative to those of the original FCal.  Irradiation tests carried out at IHEP/Protvino confirmed that at beam intensities of up to $10^{10}$ protons/s, the peak height of the ionization signal current for the case of the standard-design FCal will suffer, whereas in the case of a prototype constructed with the same basic design but with LAr gaps approximately 44\% the original size (a reduction from $269\,\mu$m to $119\,\mu$m), the peak height was shown to be stable even up to the highest beam intensities recorded in the test setup \cite{atlas-lar-perf-high-rates}.  The high-voltage sagging effects would be countered by employing protection resistors in the sFCal a factor of ten smaller than the 1-$2\,$M$\Omega$ resistors used in the FCal.

\begin{figure}[h]
\centering
\includegraphics[width=3.0in]{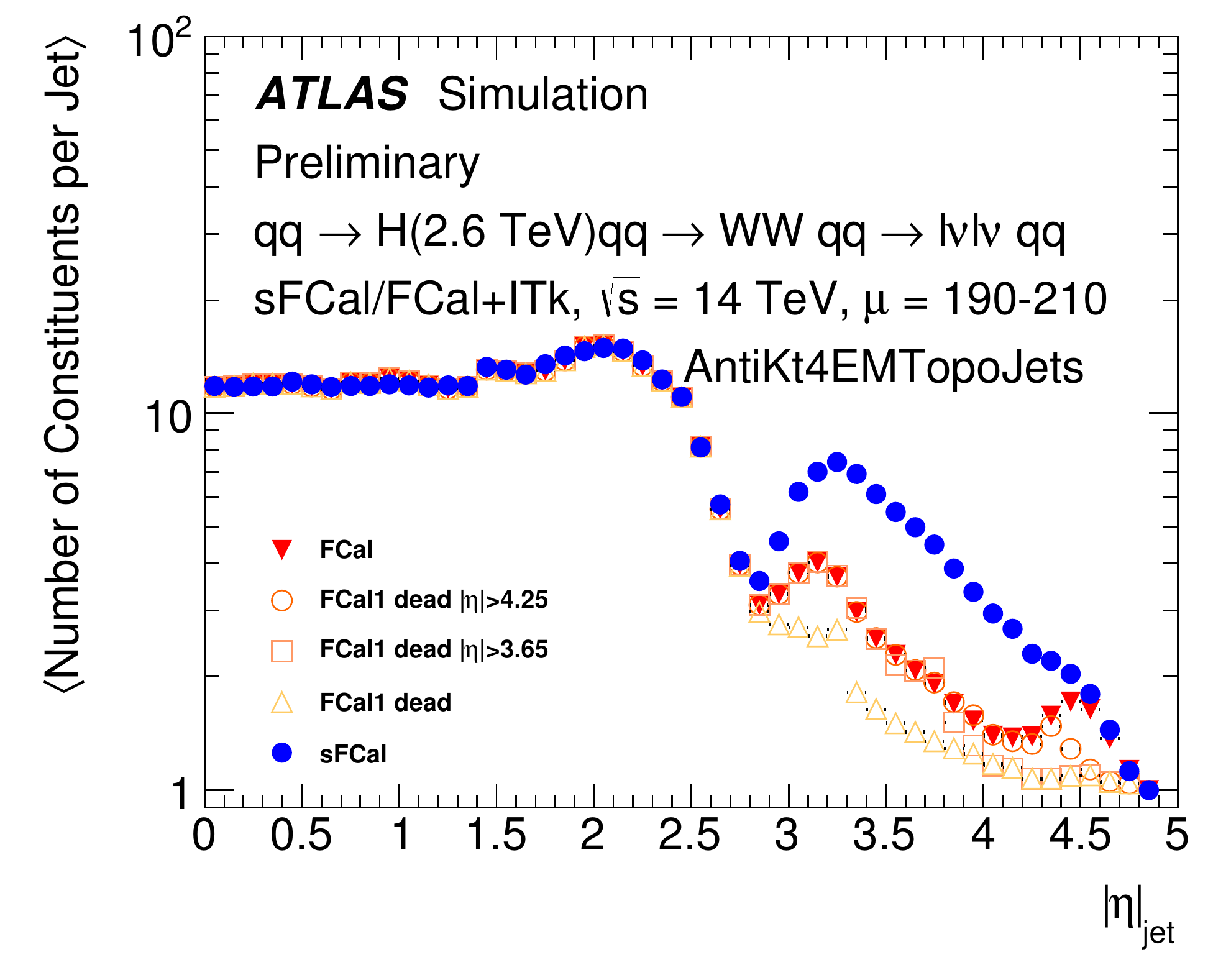}
\caption{The average number of constituents for reconstructed jets as a function of pseudorapidity as evaluated from high pile-up simulated events.  Results are shown for the proposed higher-granularity FCal (blue markers) and the standard FCal design, and in the latter both for the undegraded (red markers) and three degraded scenarios (various open orange markers).}
\label{fig:fcal-nclusters}
\end{figure}

In addition to the reduced LAr gap size, the granularity for a large fraction of readout channels in sFCal1 would be increased by a factor four compared to the original FCal1 design.  On an individual readout channel level, this increase in granularity was shown to increase the cell significance or signal-to-noise ratio in simulation.  One consequence of this improvement is an increased number of reconstructed energy clusters in the forward region.  Since the energy clusters form the base constituents of reconstructed hadronic jets, this increased granularity would also lead to an improvement in jet substructure -- previously a limitation for forward-jet substructure studies within ATLAS.  The increased jet substructure provided by the sFCal can be clearly seen in Figure~\ref{fig:fcal-nclusters} which shows the results based on the simulation studies to be described in the following section.

\begin{figure}[h]
\centering
\includegraphics[width=3.25in]{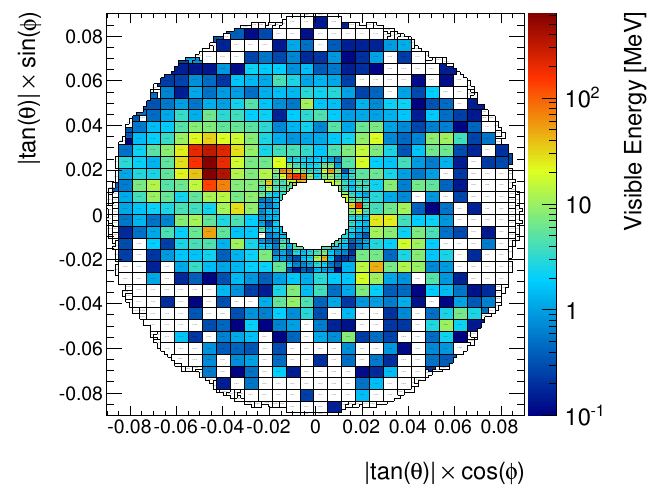}
\caption{A simulated event display depicting the visible energy deposits from a single jet in FCal1 based on the standard FCal geometry \cite{atlas-phase2-scoping-document}.}
\label{fig:fcal-jet}
\end{figure}

\begin{figure}[h]
\centering
\includegraphics[width=3.25in]{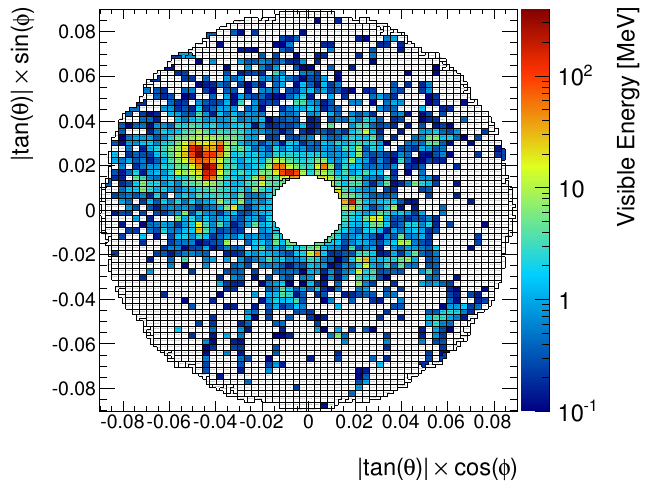}
\caption{A simulated event display depicting visible energy deposits in FCal1 for the same single jet displayed in Figure~\ref{fig:fcal-jet} but with the higher-granularity sFCal geometry \cite{atlas-phase2-scoping-document}.}
\label{fig:sfcal-jet}
\end{figure}

The improvements offered to jet substructure as a result the increased sFCal granularity can also be observed qualitatively by contrasting the two-dimensional distributions of energy deposits in the first FCal layer from a single jet in Figures~\ref{fig:fcal-jet} and \ref{fig:sfcal-jet}, which correspond to the standard FCal geometry and the higher-granularity sFCal geometry, respectively.

\subsection{Physics Studies with the sFCal Based on Simulation}\label{sec:fcal-sfcal-physics}

Dedicated studies were carried out using simulated \mbox{$\rts=14\,\TeV$} events featuring VBF-produced heavy scalar particles \mbox{($m=2.6\,\TeV)$}.  A key feature of such events is the presence of forward quark-initiated jets, typically in the FCal.  Simulated datasets were produced, with varying levels of pile-up, for both the original FCal and the improved sFCal in order to quantify the expected performance gains.  Samples featuring the standard FCal geometry but with varying levels of degradation were also produced for comparison.   Many of these studies involved pile-up suppression techniques with a specific focus on jets in the forward region.  One common strategy in such studies was to use some form of jet area-based $\pT$ subtraction in order to initially suppress the contribution from pile-up jets \cite{atlas-pu-suppression,jet-area-correction}.

\begin{figure}[h]
\centering
\includegraphics[width=3.0in]{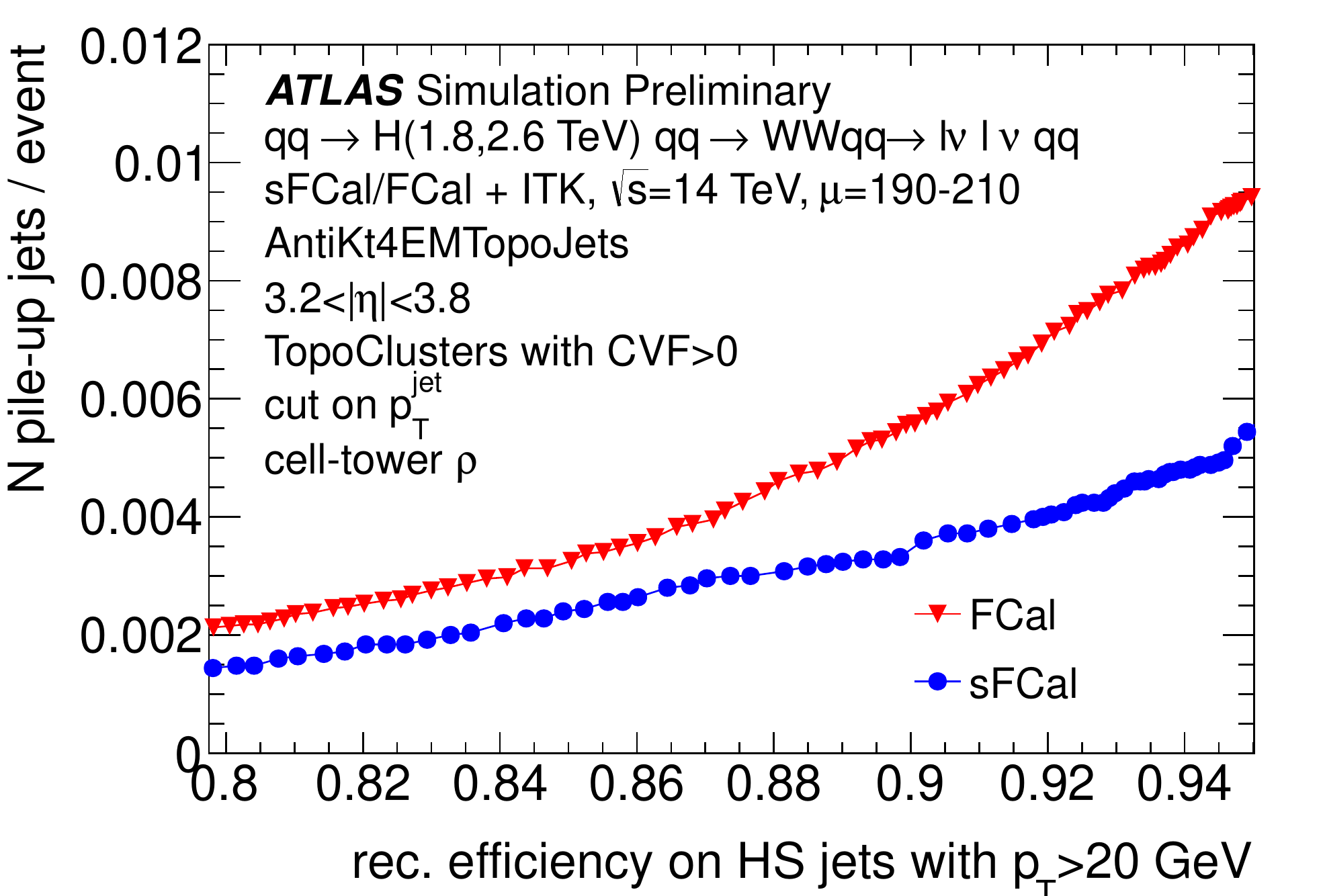}
\caption{A ROC curve showing the number of reconstructed pile-up jets per event as a function of the efficiency for identifying hard-scatter jets in heavy-scalar VBF events at \mbox{$\rts=14\,\TeV$} and with \mbox{$190<\aipbc<210$}.  Reconstructed jets were required to satisfy \mbox{$\pT>20\,\GeV$} and lie in the pseudorapidity range \mbox{$3.2<|\eta|<3.8$}.  An area-based jet \pT \ correction was arrived based on an average \pT \ density ($\rho$), evaluated from positive-energy cell towers.}
\label{fig:fcal-roc-pusuppression}
\end{figure}

In some studies the improvements offered by the sFCal were combined with tracking information from the planned Phase-II upgrade to the ATLAS inner tracker (ITk) \cite{atlas-phase2-scoping-document}.  Figure~\ref{fig:fcal-roc-pusuppression} compares, for example, the average number of reconstructed pile-up jets per event as a function of the reconstructed hard-scatter jet efficiency.  The jet classification (hard-scatter versus pile-up) is performed using a geometrical matching to truth particle jets available from the Monte Carlo simulation.  One can note from the results shown in Figure~\ref{fig:fcal-roc-pusuppression}, for example, that for a hard-scatter jet efficiency of 90\%, the number of pile-up jets using the sFCal geometry is reduced by a factor of roughly 1.7 compared with the case of the standard FCal geometry.  

This gain in performance is achieved in part by making a cut to the cluster vertex fraction (CVF), a quantity which relies on the tracks of charged particles provided by the ITk, and is a measure of the probability that a given reconstructed cluster is associated with energy deposits from hard-scatter particles.  For this study a tracking coverage of $|\eta|<4$ and an idealized tracking detector resolution were assumed.  The result nevertheless demonstrates the clear advantage of the finer granularity sFCal in suppressing pile-up.

The reconstructed jet four-vectors in all simulation studies were formed by sequentially grouping together positive-energy topological clusters using the anti-$k_{T}$ algorithm with a radius parameter $R=0.4$ \cite{antikt}.

\subsection{Replacement Decision}\label{sec:fcal-decision}

An exhaustive analysis was carried out to assess each of the risks associated with the potential removal of the current FCal modules and an installation of their proposed sFCal replacements.  This included a comprehensive and detailed study of all engineering aspects associated with the construction of the sFCal modules themselves, the FCal removal and sFCal insertion, and any additional cabling.  Due to the position of the current FCal within the sealed end-cap cryostat, the sFCal upgrade would necessarily involve opening the cryostat after several years of LHC running -- a process which would require the removal of some welds at the cryostat openings.  An associated risk involved thermal and mechanical stresses on some of the more fragile detector components and the potential creation of new electric shorts from the infall of debris which could lead to additional permanentaly dead channels.  Such actions were felt to pose additional unwanted risks to the electromagnetic and hadronic end-cap calorimeters which share the same cryostat.

Overall the risks of replacement were deemed to outweigh any potential gains in physics performance offered by the sFCal.  Given the results of studies which suggest the FCal should continue to function throughout the \mbox{HL-LHC} period, albeit at a degraded but tolerable level, a recent decision was made by the collaboration that ATLAS will continue to operate with the original FCal in place.

Future work by the LAr community will focus on the ability to anticipate and understand, as well as to correct for, any deterioration in the FCal signal pulses at these higher instantaneous luminosities.


\section{High-Granularity Timing Detectors}\label{sec:hgtd}

The construction of a high-granularity timing detector (HGTD) is currently being considered for the ATLAS Phase-II upgrades \cite{atlas-phase2-scoping-document}.  The basic principle behind the HGTD involves exploiting the time-of-flight information of particles in order to aid in suppressing contributions from additional pile-up vertices.  This is possible by first taking advantage of a scheme by which the orientation of the colliding proton bunches can be manipulated by the LHC \cite{lhc-crab-kissing}.  The success of the HGTD relies in part on the ability to achieve timing resolution on the order of a few tens of picoseconds.

\subsection{Proposed Design}\label{sec:hgtd-design}

The leading design is based on a silicon-based low-gain avalanche detector (LGAD) \cite{lgad} and features, in the case of the baseline option, four active silicon (Si) layers each approximately \mbox{$150\,\mu$m} thick (HGTD-Si option).  Unlike the proposed sFCal replacement, the HGTD would be situated between the barrel and end-cap cryostats and within an envelope of \mbox{$\Delta z<60\,$mm}.  The active area of the HGTD would cover the pseudorapidity range \mbox{$2.4\leq|\eta|\leq4.2$}, corresponding to a radial separation \mbox{$90\,$mm$\,\leq R \leq600\,$mm} as measured from the LHC beampipe.  A variation on this baseline option is considered in which three interleaved tungsten absorbers are added between the active Si layers in order to act as a preshower detector, allowing for $\gamma/\pi^{0}$ conversions (HGTD-SiW option).  The added tungsten layers would cover only the pseudorapidity range \mbox{$2.4\leq|\eta|\leq3.2$}, thereby limiting any detrimental effects on the FCal energy resolution.

\begin{figure}[h]
\centering
\includegraphics[width=3.5in]{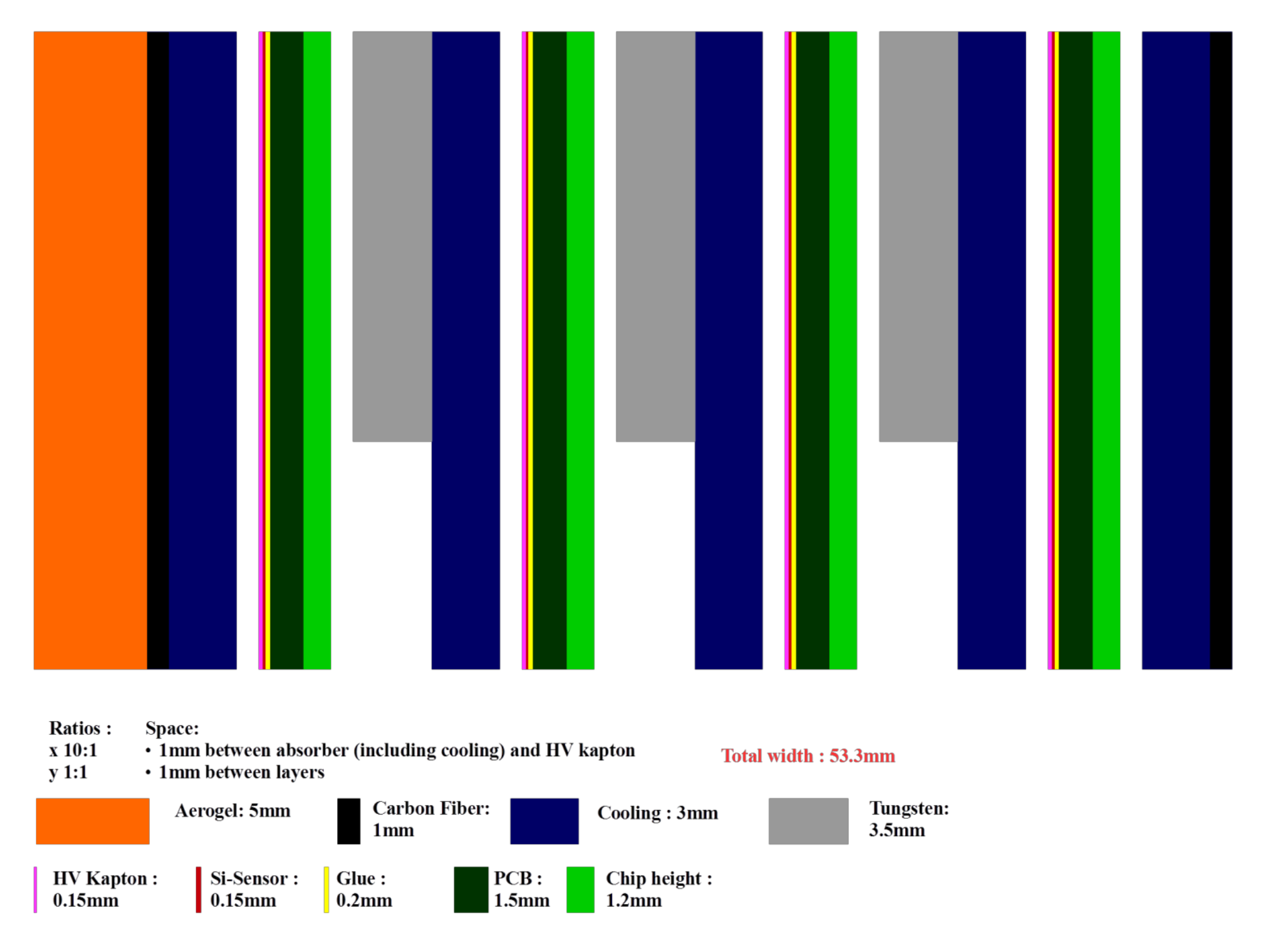}
\caption{A schematic view of the HGTD-SiW in positive-z and R. The thicknesses in z of the various materials as well as the material types employed in the simulation are listed in the legend. The tungsten absorber begins at a radial separation of \mbox{$R=285\,$mm} as measured from the LHC beam axis.}
\label{fig:hgtd-layout-side}
\end{figure}

\begin{figure}[h]
\centering
\includegraphics[width=3.0in]{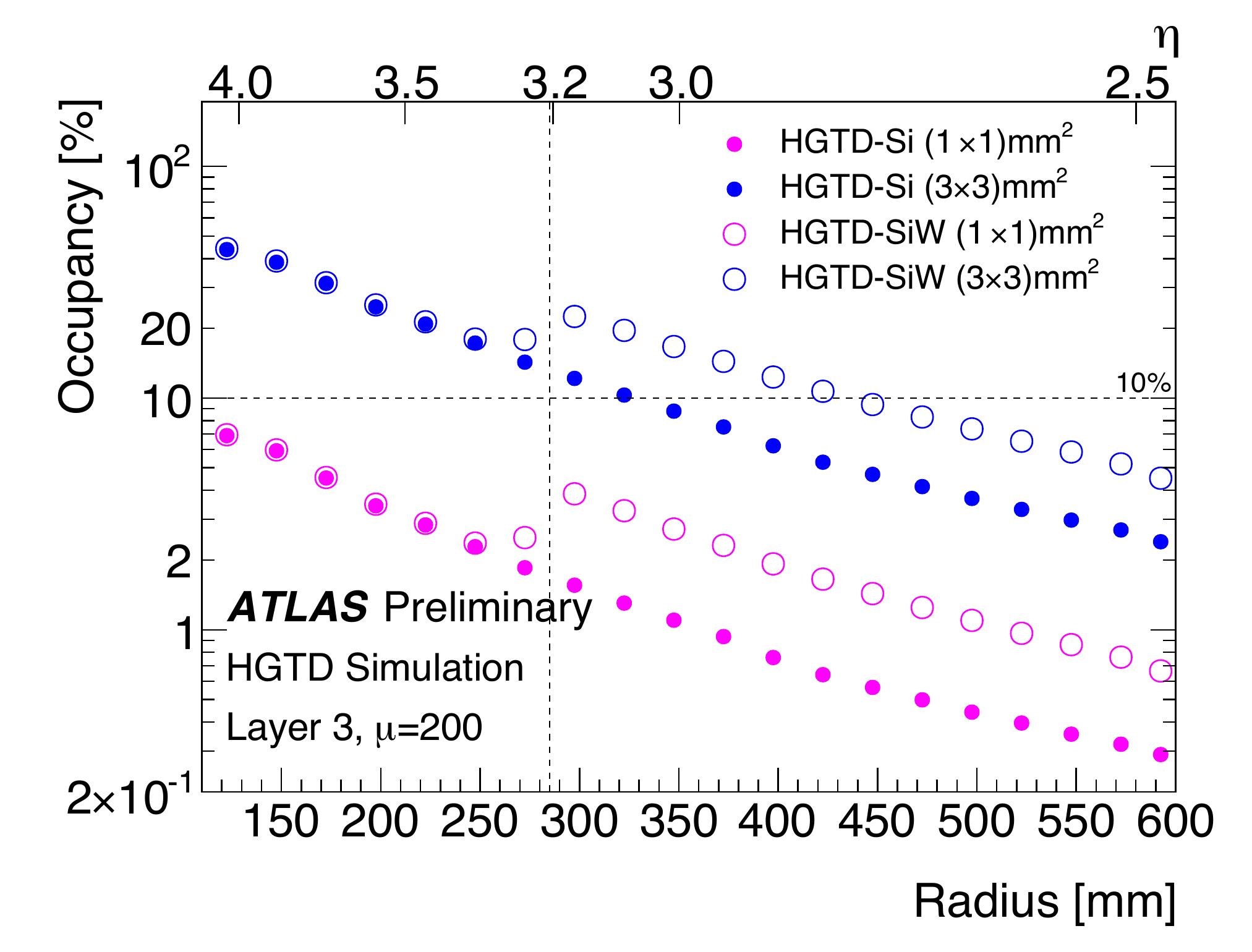}
\caption{The occupancy of the HGTD plotted as function of the radius for a pileup of $\mu=200$ based on simulation, shown separately for granularities of (1$\times$1) mm$^2$ and (3$\times$3) mm$^2$ in the third sampling layer.  The presence of the tungsten absorbers beginning at a radius of \mbox{$285\,$mm} lead to a higher occupancy for the HGTD-SiW compared to the HGTD-Si.}
\label{fig:hgtd-occupancy-simulation}
\end{figure}

\begin{figure}[h]
\centering
\includegraphics[width=2.5in]{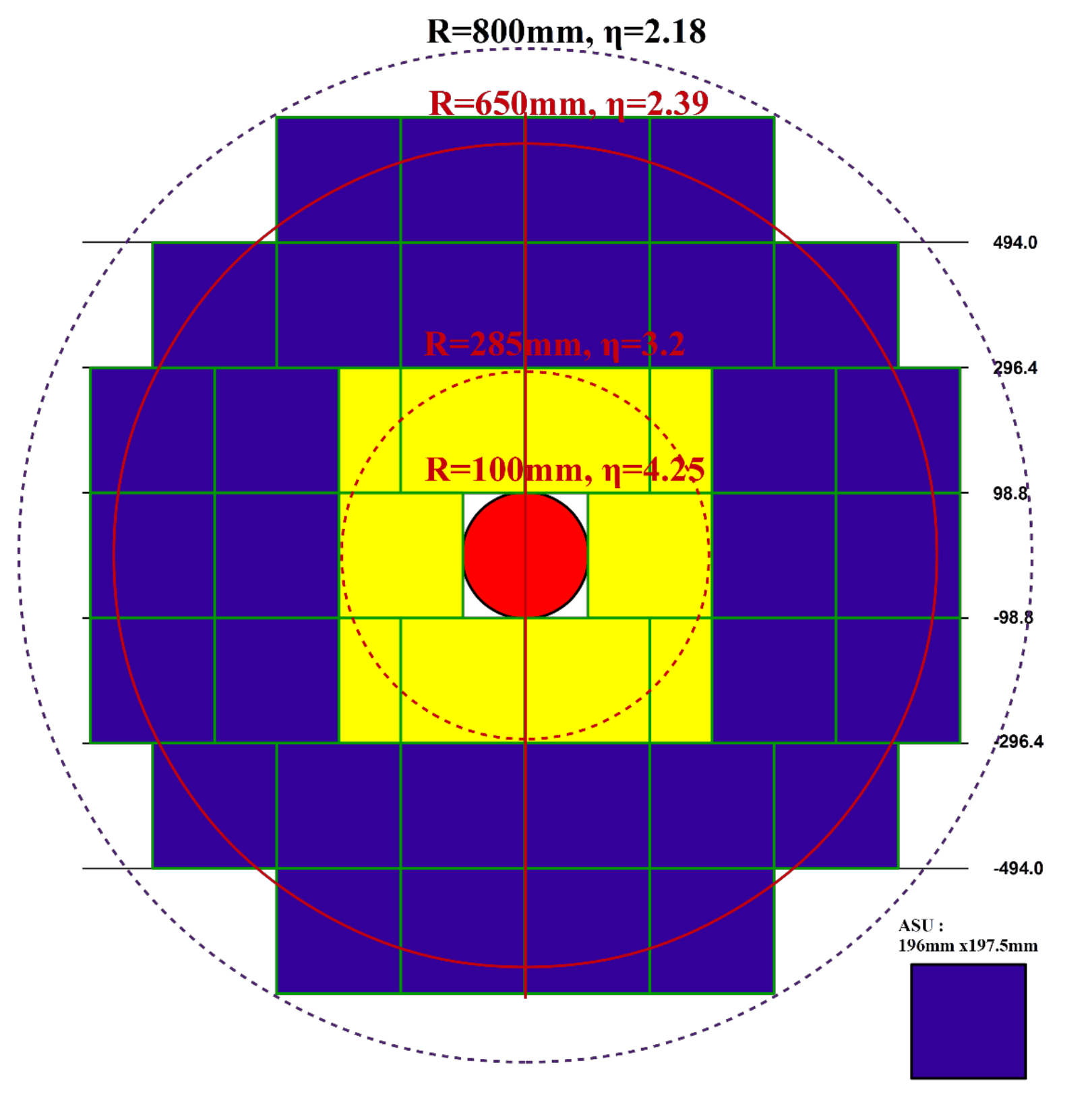}
\caption{Schematic view of the transverse plane of a HGTD layer. The yellow regions have a granularity of \mbox{$(1\times1)\,$mm$^2$}, the blue regions \mbox{$(3\times3)\,$mm$^2$}. The green lines mark the border of an ASU.}
\label{fig:hgtd-layout-front}
\end{figure}

The choice of final granularity for each individual active sensor unit (ASU) was based on the ability to have an occupancy at or below the 10\% level at high pile-up levels ($\aipbc\sim200$).  As can be seen in Figure~\ref{fig:hgtd-occupancy-simulation}, enforcing such a requirement allows an ASU to have a granularity of \mbox{$(3\times3)\,$mm$^2$} for the outer radii \mbox{($R\geq300\,$mm)}, but requires a finer granularity of \mbox{$(1\times1)\,$mm$^2$} in the inner region based on dedicated simulated studies.  A schematic view of one of the individual HGTD layers is shown in Figure~\ref{fig:hgtd-layout-front}.  The anticipated total number of individual readout channels for the HGTD is approximately \mbox{$3.4\times\,10^6$}.

\subsection{Beam \& Irradiation Tests}\label{sec:hgtd-beam-tests}

\begin{figure}[h]
\centering
\includegraphics[width=3.0in]{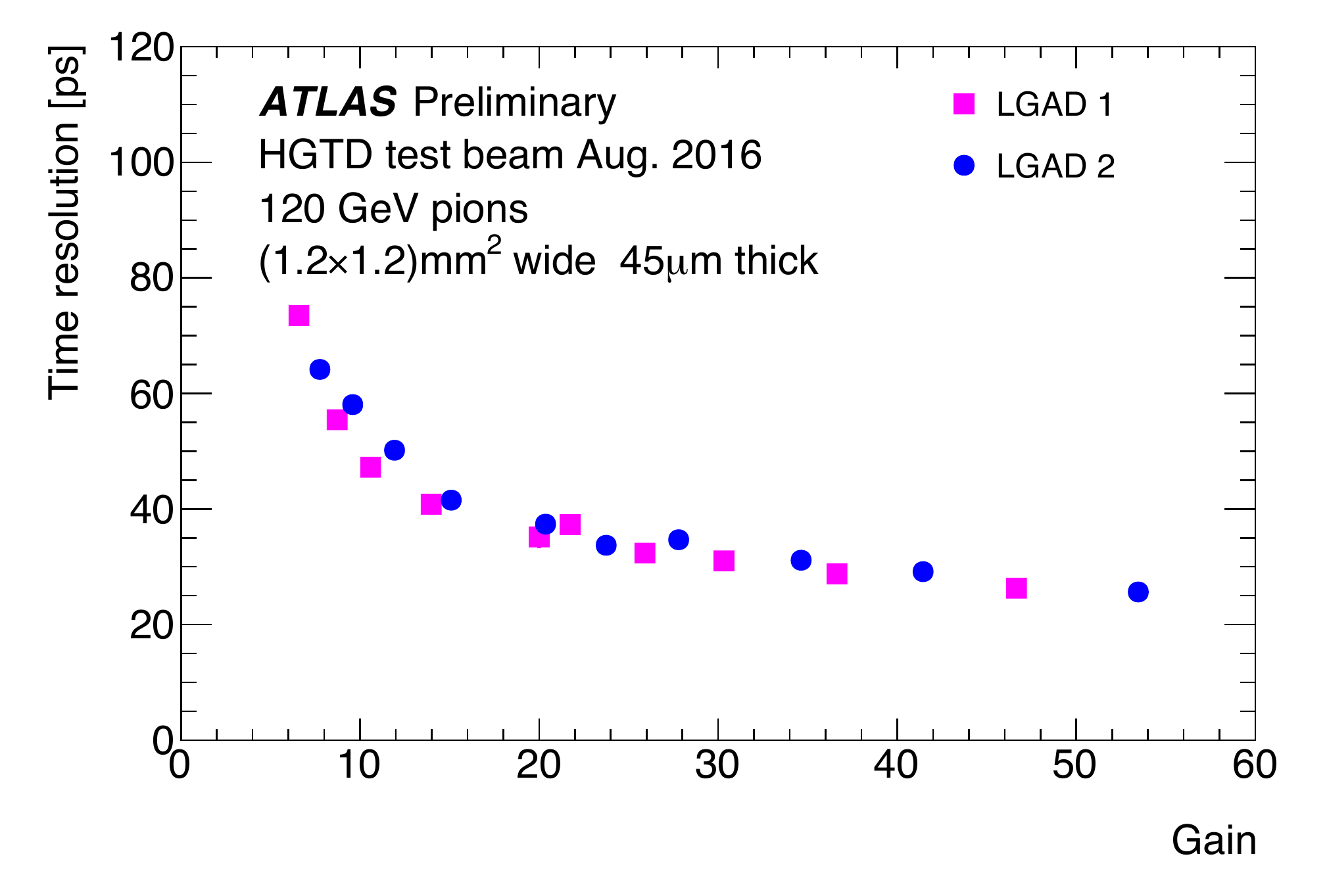}
\caption{Measured time resolution as a function of gain based on the results of the beam test as described in the text.}
\label{fig:hgtd-timereso}
\end{figure}

Two single-pad LGAD sensors with an area of \mbox{1.2 $\times$ 1.2 mm$^2$} and approximate width of \mbox{$45\,\mu$m} were exposed to \mbox{$120\,$\GeV} protons from the SPS H6B beam line at CERN in August 2016.  The total integrated charge and timing resolution were measured as a function of applied bias voltage and gain.  Figure~\ref{fig:hgtd-timereso} shows the measured timing resolution versus the gain, demonstrating that for a high-gain scenario a timing resolution on the order of \mbox{$\sigma_t= 26\,$ps} can be achieved.

A further requirement for the HGTD is that it be able to perform well in a high-radiation environment such as that expected in the forward region of ATLAS during the \mbox{HL-LHC}.  A series of irradiation tests have demonstrated that a gain of $>5$ is achievable with an applied bias voltage of \mbox{$750\,$V} following neutron irradiation of approximately \mbox{$4 \times 10^{15}\,$n$_{\textrm{eq}}$/cm$^2$}.  

Should the HGTD be endorsed by the collaboration, the final design and layout as well as hardware components will be governed by the results of these as well as future studies carried out over the coming few years. 

\section{Conclusion}\label{sec:conclusion}
The harsh conditions leading up to and beyond the start of the \mbox{HL-LHC} will require a number of hardware upgrades to the ATLAS LAr calorimeter system.  The Phase-I and Phase-II upgrades will help to maximize the discovery potential in ATLAS for signal events featuring large energy deposits in the calorimeter or \MET \ signatures.

The incorporation of Super Cells during Phase-I will increase the input granularity to the L1 trigger up to a factor of 10, allowing several discriminating shower-shape variables to be exploited in order to suppress the background rates while retaining high signal efficiencies.  The subsequent replacement of the back-end electronics during Phase-II will allow the full calorimeter granularity to be incorporated at trigger level.

Following a careful analysis of potential physics gains and associated risks, a decision was made not to replace the current FCal with an upgraded sFCal during Phase-II.  Focus will instead be placed on correcting for any potential degradation in the FCal signals.

Significant progress has been made in the context of the proposed HGTD, one of the major LAr Phase-II upgrade projects.  Full-simulation datasets incorporating the HGTD are available and dedicated analyses are ongoing.  In terms of the hardware, early results from beam and irradiation tests have served as a proof of concept for the device.  Further such tests and simulation studies are planned in the near future.

\ifCLASSOPTIONcaptionsoff
  \newpage
\fi

\end{document}